\documentclass[10pt,conference]{IEEEtran}
\usepackage[backend=biber, style=ieee]{biblatex}
\usepackage{amsmath}
\usepackage{amssymb}
\usepackage{graphicx}
\usepackage{subcaption}
\usepackage{algorithm}
\usepackage{algpseudocode}
\usepackage{booktabs}
\usepackage{placeins}
\usepackage{xcolor}

\usepackage{tikz}
\usetikzlibrary{shapes.geometric, calc, positioning, arrows.meta, backgrounds, fit}

\definecolor{CircuitABlue}{RGB}{66,133,244} 
\definecolor{CircuitBRed}{RGB}{234,67,53}  
\definecolor{PaddingGray}{RGB}{200,200,200}  
\definecolor{CotenantGreen}{RGB}{52,168,83}  
\definecolor{CotenantRed}{RGB}{234,67,53}

\definecolor{SetupGray}{HTML}{F2F2F2}
\definecolor{BaselineBlue}{HTML}{D9EAF7}
\definecolor{StressGreen}{HTML}{EAF7D9}
\definecolor{MaxStressRed}{HTML}{F7D9D9}

\DeclareMathOperator*{\argmin}{arg\,min}

\author{\IEEEauthorblockN{Andrew Woods}
\IEEEauthorblockA{MU Quantum Innovation Center\\
Electrical Engineering and Computer Science Department\\
University of Missouri\\
Columbia, MO, USA}
 \and
\IEEEauthorblockN{Chi-Ren Shyu}
\IEEEauthorblockA{MU Quantum Innovation Center\\
Electrical Engineering and Computer Science Department\\
University of Missouri\\
Columbia, MO, USA}
}

\title{Toward Secure Multitenant Quantum Computing: Circuit Affinity, Crosstalk Patterns, and Grouping Strategies}

\begin{document}
\maketitle

\begin{abstract}
  Multitenancy increases throughput and reduces costs in cloud-based quantum computing, but concurrent job execution introduces security risks through inter-circuit crosstalk. We characterize the structural predictability of these interference patterns across seven IBM superconducting processors, spanning Heron (r1-r3) and Nighthawk (r1) architectures and five different circuit types. We evaluate pairwise interactions, by applying the Structural Similarity Index (SSIM) and a structural $t$-statistic to the concurrent execution of five foundational quantum circuits (QAOA, Grover’s, QPE, QFT, and ZZFeatureMap), we quantify behavioral consistency across disparate hardware. Our results identify three types of circuits: universally aggressive, universally sensitive, and cotenant-dependent circuits. Aggressive circuits, such as Grover’s Algorithm, exhibit a statistically significant interference pattern, yielding a $t$-statistic range of $[1.37,2.61]$ relative to the standalone baselines across all tested pairings. Conversely, sensitive circuits, such as the Quantum Fourier Transform, demonstrate a disproportionate susceptibility to multitenant execution, showing high deviations from single-tenant computational behavior. We demonstrate that crosstalk signatures are highly consistent within architectural revisions—with intra-revision similarity reaching $0.77$ (Hr3) and $0.68$ (Hr2)-while inter-revision similarity drops to $0.43$. Furthermore, we identify a ``topological decoupling" between Heavy-Hex and square lattice systems, where structural similarity falls to $0.01$ between Heron r1 and Nighthawk r1. These findings provide an empirical foundation for hardware-aware schedulers to strategically pair jobs, maximizing system utilization while preserving computational integrity.
\end{abstract}

\section{Introduction}

Quantum computers are an expensive resource with an ever-growing demand and increasing industrial and academic interest from fields such as chemistry, material science, life sciences, and finance \cite{McKinseyQuantum2024}. As quantum circuits increase in qubit count, the circuit complexity also increases. Current noisy intermediate scale quantum (NISQ) computers have limitations on their processing capacity. To combat this, users design smaller circuits to improve computational reliability, resulting in underutilization of the quantum computer. As quantum computing evolves as a computational paradigm and demand for quantum computing as a service (QCaaS) increases, queue lengths and wait times increase proportionally.

Multitenancy, a technique which allows multiple users to concurrently access shared hardware, improves hardware utilization~\cite{qVMs}. However, this paradigm introduces security and integrity risks~\cite{ToAntiVirus,DesignAntiVirus, Hawley2024, crosstalk_sidechannel, shuttle_exploit, muzzle_shuttle, Maurya_2024, ovaskainen2025quantumsoftwaresecuritychallenges}. This work focuses on crosstalk, which has been shown as an attack vector for fault-injection~\cite{ToAntiVirus, DesignAntiVirus, Hawley2024} as well as a side-channel for leaking information~\cite{crosstalk_sidechannel, Maurya_2024}.

Previous works have shown that specific circuit designs can be developed to reduce the integrity of quantum computation by producing excessive amounts of crosstalk through high-crosstalk operations~\cite{ToAntiVirus, DesignAntiVirus}. Furthermore, other studies have shown that these crosstalk attacks can be placed into common algorithmic circuits in a quantum circuit library~\cite{Hawley2024}, causing users to unknowingly induce interference in neighboring workloads. Beyond crosstalk as a vector for fault-injection, works like~\cite{crosstalk_sidechannel, Maurya_2024, noise_security_impact} demonstrated that crosstalk can be used as a side-channel vector to deduce the structure and qubit state of the victim circuit. 

While these works show vulnerabilities based on intentional misuse of crosstalk, even in the context of benign multitenancy the computational integrity suffers. Thus, realizing viable multitenancy requires a granular understanding of inter-circuit interference. In this work, we empirically characterize the pairwise compatibility of quantum circuits to identify stable interference signatures across diverse hardware generations. This approach can guide a scheduler that seeks to minimize the interference between circuits in a cotenant scenario. This provides an empirical foundation for making multitenant processors more secure in terms of crosstalk attacks. Specifically, we:

\begin{itemize}
  \item Test five circuit types (Max-Cut QAOA (QAOA), Grover's Algorithm (GA), Quantum Fourier Transform (QFT), Quantum Phase Estimation (QPE), and ZZFeatureMap (ZZFM)) in a pairwise concurrent execution to get an interference profile between circuits across seven quantum machines and quantum architectural revisions (from Heron to Nighthawk).
  \item Quantify intra-generational processor interference patterns to determine the consistency of crosstalk patterns between quantum processor generations.
  \item Identify universally aggressive circuits, universally sensitive circuits, and cotenant-dependent circuits.
  \item Propose a scheduling optimization problem that can reduce the interference in multitenant scenarios.
\end{itemize}

\section{Background}

\subsection{Crosstalk}
Crosstalk refers to the unintended interactions between qubits, frequently arising from signal leakage between driving pulses. Crosstalk can be further defined as the violation of locality and independence of operations on the processor \cite{Sarovar2020detectingcrosstalk}. Crosstalk results in reduced integrity and leads to side-channel and fault-injection attacks in multitenant settings. 

\subsection{Hardware Platforms}

The processors used for evaluation in this work span four architectural revisions of superconducting processors from Heron r1, r2, and r3 and Nighthawk r1. These devices use tunable couplers that defines the processors topology. Heron processors utilize the Heavy-Hex lattice topology to prioritize qubit isolation and Nighthawk makes use of the square-lattice topology. This difference in topology results in behavioral differences in crosstalk.

\subsection{Structural Similarity Index Measure}
We evaluate crosstalk behavior between pairs of circuits and the behavioral differences between a few types of processors. The method of choice to compare processor behavioral similarity is the structural similarity index measure (SSIM) \cite{SSIMPaper}. In general, the SSIM is used for identifying the similarity between two images, and the outputs range from -1 to 1. The structural similarity between two random variables $x$ and $y$ is measured as
\begin{equation}
\text{SSIM}(x, y) = \frac{\left(2 \mu_x \mu_y + c_1\right) \left(2 \sigma_{xy} + c_2\right)}{\left(\mu_x^2 + \mu_y^2 + c_1\right) \left(\sigma_x^2 + \sigma_y^2 + c_2\right)},
\end{equation}
where $\mu_x, \mu_y$ are the means of $x$ and $y$ respectively, $\sigma_x^2$, $\sigma_y^2$ are variances and $\sigma_{xy}$ is the sample covariance of $x, y$. $c_1$ and $c_2$ are variables to stabilize the division on a weak denominator. Typically, $c_1 = (k_1L)^2$ and $c_2 = (k_2L)^2$ where $L$ is the dynamic range of the data and $k_1 = 0.01, k_2 = 0.03$. This work utilizes interference matrices between the circuits as the images of evaluation, where a pixel refers to the $t$-statistic interference between the two states.

\section{Methods}
\label{sec:method}

\subsection{Circuit Selection}

\begin{table}[htbp]
\caption{Structural Characteristics of Benchmarked Quantum Circuits}
\label{table:circuit_characteristics}
\centering
\begin{tabular}{lcccc}
\toprule
\textbf{Circuit} & \textbf{Avg. Depth} & \textbf{Depth Range} & \textbf{Avg. 2Q Gates} & \textbf{2Q Ratio} \\
 & ($D_{avg}$) & [$D_{min}, D_{max}$] & ($N_{2Q}$) & (\%) \\
\midrule
GA & 1746.53 & [1317, 2333] & 619.88 & 22.30\%  \\
QAOA   & 183.60  & [153, 222]   & 62.94  & 19.72\%  \\
QPE  & 150.21  & [131, 166]   & 51.13  & 22.38\%  \\
ZZFM  & 236.49  & [191, 274]   & 79.76  & 22.00\%  \\
QFT    & 117.13  & [92, 136]    & 41.36  & 23.85\%  \\
\bottomrule
\addlinespace[1ex]
\multicolumn{5}{l}{\textsuperscript{*}Values represent averages across all test iterations.}
\end{tabular}

\end{table}

Testing the impact of multicircuit execution is resource-heavy, because it requires significant compute resources to execute the full suite of experiments. We select five circuits that represent functional workloads. The circuits exhibit depths between 118 layers for QFT to 1747 layers for Grover's Algorithm. By evaluating this diverse set of quantum circuits, we can characterize how different algorithmic motifs influence crosstalk signatures in shared hardware environments:

\begin{itemize}
    \item Max-cut quantum approximate optimization algorithm (QAOA) \cite{farhiQAOA}
    \item Grover's algorithm (GA) \cite{grover1996}
    \item Quantum Fourier transform (QFT) \cite{coppersmithQFT}
    \item ZZFeatureMap (ZZFM) 
    \item Quantum phase estimation (QPE) \cite{kitaev}
\end{itemize}

\begin{table}[htbp]
  \centering
  \caption{Summary of Quantum Processors and Experimental Volume}
  \label{tab:machines}
  \begin{tabular}{@{}llcc@{}}
  \toprule
  \textbf{Backend Name} & \textbf{Architecture} & \textbf{Qubits} & \textbf{Trials ($n$)} \\ \midrule
  \texttt{ibm\_torino}    & Heron r1             & 133             & 16                  \\
  \texttt{ibm\_fez}       & Heron r2             & 156             & 24                  \\
  \texttt{ibm\_marrakesh} & Heron r2             & 156             & 24                  \\
  \texttt{ibm\_kingston}  & Heron r2             & 156             & 24                  \\
  \texttt{ibm\_pittsburgh}& Heron r3             & 156             & 24                  \\
  \texttt{ibm\_boston}    & Heron r3             & 156             & 24                  \\
  \texttt{ibm\_miami}     & Nighthawk r1         & 120             & 2                   \\ \midrule
  \textbf{Total}          &                      &                 & \textbf{138}        \\ \bottomrule
  \end{tabular}
\end{table}

Table \ref{table:circuit_characteristics} details the average depth and 2-qubit gate count of each circuit types.

\subsection{Hardware Distribution}

The data was collected on seven IBM quantum machines, summarized in Table \ref{tab:machines}. We selected these machines to leverage topological similarities and differences as experimental controls. By evaluating multiple machines with the same architecture (e.g., Heron R2), we establish a baseline for inter-hardware consistency. The inclusion of the Nighthawk processor allows us to evaluate the impact of topological impact on cotenant affinity between the selected circuits.

\subsection{Design of Experiment}

\begin{algorithm}[htbp]
  \caption{Circuit Mapping Selection}
  \label{alg:circuit_map_selection}
  \begin{algorithmic}[1]
      \Procedure{GetMapsForCircuits}{$\mathcal{C}, \mathcal{T}, p$}
          \Require{$\mathcal{C}$: list of circuits; $\mathcal{T}$: coupling map; $p$: padding depth.}
          \State $\mathcal{M} \gets [ \: ]$ \Comment{Resulting set of mappings $M_i$}
          \State $\mathcal{R} \gets \emptyset$ \Comment{Globally reserved qubit set}
          \State $\mathcal{T}' \gets \mathcal{T}$ \Comment{Residual subgraph of available qubits}
          \State $\mathcal{F} \gets \mathcal{T}$ \Comment{Frontier set for potential mapping seeds}
          
          \For{$C \in \mathcal{C}$}
            \State $q \gets \text{Random}(\mathcal{T}')$ \Comment{Select initial seed qubit}
            \State $M' \gets \text{BFS\_set}(q, \mathcal{T}', \text{size}(C))$ \Statex \hspace{\algorithmicindent} \Comment{Identify contiguous subset}
            
            \While{$M' = \perp$} 
            \Statex \hspace{\algorithmicindent} \Comment{Retry mapping if contiguous fit fails}
              \State $q \gets \text{Random}(\mathcal{F})$
              \State $M' \gets \text{BFS}(q, \mathcal{T}')$
            \EndWhile
            
            \State $P \gets \text{BFS\_pad}(\mathcal{T}'\text{.neighbors}(M'), \mathcal{T}', p)$ 
            \Statex \hspace{\algorithmicindent} \Comment{Construct safety buffer of depth $p$}
            
            \State $\mathcal{T}' \gets \mathcal{T}' \setminus (M' \cup P)$ \Comment{Update residual graph}
            \State $\mathcal{R} \gets \mathcal{R} \cup M' \cup P$ \Comment{Update reserved set}
            \State $\mathcal{F} \gets \mathcal{T}\text{.neighbors}(\mathcal{R})$ \Comment{Recompute frontier}
            \State $M \gets \text{select\_qubit\_mapping}(C, M')$ 
            \Statex \hspace{\algorithmicindent} \Comment{Finalize layout}
            \State $\mathcal{M}\text{.append}(M)$
          \EndFor
          \State \Return $\mathcal{M}$
      \EndProcedure
  \end{algorithmic}
\end{algorithm}

\begin{algorithm}[htbp]
  \caption{Sandwiched Baseline-Cotenant Calibration}
  \label{alg:circuit_matching}
  \begin{algorithmic}[1]
    \Procedure{SandwichedExecution}{$\mathcal{C}$, $\mathcal{Q}$, $p$}
      \Require{$\mathcal{C}$: list of circuits; $\mathcal{Q}$: target hardware; $p$: padding depth.}
      \State $\mathcal{M} \gets \text{GetMapsForCircuits}([\mathcal{C}[0], \mathcal{C}[1]], \mathcal{Q}.\mathcal{T}, p)$
      \State $M_0, M_1 \gets \mathcal{M}[0], \mathcal{M}[1]$ \Comment{Static mappings for $C_A, C_B$}
      \State $\mathcal{J} \gets [\: ]$ \Comment{Global job execution queue}
      \State $\mathcal{V} \gets [\: ]$ \Comment{Set of baseline Solo transpiled circuits}
      
      \Statex \texttt{// Phase 1: Baseline (Solo) Generation}
      \For{$c_0 \in \mathcal{C}$}
        \State $T_0 \gets \text{Transpile}(c_0, \mathcal{Q}, M_0, \text{seed}=42,$
        \Statex \hspace{\algorithmicindent} \text{optimization\_level}=2) \Comment{Force $M_0$ constraint}
        \State $\mathcal{J}\text{.append}(T_0)$
        \State $\mathcal{V}\text{.append}(T_0)$
      \EndFor

      \Statex \texttt{// Phase 2: Cotenant Stress (Zipped)}
      \For{$T_0 \in \mathcal{V}$}
        \For{$c_1 \in \mathcal{C}$}
          \State $T_1 \gets \text{Transpile}(c_1, \mathcal{Q}, M_1, \text{seed}=42,$
          \Statex \hspace{\algorithmicindent\algorithmicindent} \text{optimization\_level}=2)
          \State $T \gets \text{transpiled\_zip}(T_0, T_1)$ \Comment{Apply Eq. \ref{eq:zipped}}
          \State $\mathcal{J}\text{.append}(T)$
        \EndFor
      \EndFor

      \Statex \texttt{// Phase 3: Post-Stress Baseline}
      \For{$T_0 \in \mathcal{V}$}
        \State $\mathcal{J}\text{.append}(T_0)$
      \EndFor
    \EndProcedure
  \end{algorithmic}
\end{algorithm}

\begin{figure*}[htbp]
  \centering
  \resizebox{\textwidth}{!}{
  \begin{tikzpicture}[
      node distance = 1.0cm,
      every node/.style={font=\small}, 
      box/.style = {rectangle, draw=black, fill=white, minimum width=3.2cm, minimum height=0.8cm, text centered, font=\small, thick},
      process/.style = {rectangle, rounded corners, draw=black!70, fill=gray!10, minimum width=3.5cm, minimum height=0.8cm, text centered, font=\small, thick},
      qubit/.style = {circle, draw=black, fill=white, minimum size=3.5mm, inner sep=0pt},
      a_qubit/.style = {qubit, fill=CircuitABlue},
      b_qubit/.style = {qubit, fill=CircuitBRed},
      p_qubit/.style = {qubit, fill=PaddingGray, draw=PaddingGray!80!black},
      job_base/.style = {rectangle, draw=CircuitABlue!80, fill=CircuitABlue!20, minimum width=2.4cm, minimum height=0.6cm, text centered, font=\tiny, thick},
      job_zipped/.style = {rectangle, draw=CotenantRed!80, fill=CotenantRed!20, minimum width=2.4cm, minimum height=0.6cm, text centered, font=\tiny, thick},
      arrow/.style = {thick, -{Stealth[length=2mm]}, black},
      coupler/.style = {draw=gray!40, line width=2.5pt, shorten >=0.5pt, shorten <=0.5pt},
      job_base_B/.style = {rectangle, draw=CircuitBRed!80, fill=CircuitBRed!20, font=\tiny, thick, minimum width=2.4cm, minimum height=0.6cm}
  ]

      % stage 1
      \node (input1) [process] at (0, 0) {Input: Circuits $\mathcal{C} = [C_A, C_B]$, Topology $\mathcal{T}$, $p$};
      \node (start1) [box, below=0.6cm of input1, font=\large] {Stage 1: Spatial Partitioning};

      \begin{scope}[shift={(-3,-3.9)}, scale=0.5, local bounding box=first_step]
          \foreach \x in {0,1,2,3} \foreach \y in {0,1, 2} \node[qubit] (latA_\x\y) at (\x,\y) {};
          \node (labelA) at (0, -1.2) {\tiny Target $\mathcal{T}$};

                    % horizontal couplers
      \foreach \x in {0,1,2} { 
      \foreach \y in {0,1,2} {
          \draw[coupler] (latA_\x\y) -- (latA_\the\numexpr\x+1\relax\y);
      }
      }

      % vertical couplers
      \foreach \x in {0,1,2,3} {
      \foreach \y in {0,1} { 
          \draw[coupler] (latA_\x\y) -- (latA_\x\the\numexpr\y+1\relax);
      }
      }
      \end{scope}

      \node (bfs_text) [text width=2.0cm, font=\tiny, align=center] at (-0.2, -3.3) {BFS select Mapping $M_1 = M'$};

      \begin{scope}[shift={(0.8,-3.9)}, scale=0.5, local bounding box=mapping_step]
          \foreach \x in {0,1,2,3} \foreach \y in {0,1,2} \node[qubit] (latB_\x\y) at (\x,\y) {};
          \node[a_qubit] at (0,0) {}; \node[a_qubit] at (1,0) {}; 
          \node (labelB) at (1, -1.2) {\tiny Map Circuit $M_1$};

                 % horizontal couplers
      \foreach \x in {0,1,2} { 
      \foreach \y in {0,1,2} {
          \draw[coupler] (latB_\x\y) -- (latB_\the\numexpr\x+1\relax\y);
      }
      }

      % vertical couplers
      \foreach \x in {0,1,2,3} {
      \foreach \y in {0,1} { 
          \draw[coupler] (latB_\x\y) -- (latB_\x\the\numexpr\y+1\relax);
      }
      }
      \end{scope}

      \begin{scope}[shift={(0.8, -6.5)}, scale=0.5, local bounding box=padding_step]
        \foreach \x in {0,1,2,3} \foreach \y in {0,1,2} \node[qubit] (lat_padding_\x\y) at (\x, \y) {};
        \node[a_qubit] at (0,0) {}; \node[a_qubit] at (1,0) {}; 
        \node[p_qubit] at (0,1) {}; \node[p_qubit] at (1,1) {}; \node [p_qubit] at (2,0) {};

              % horizontal couplers
      \foreach \x in {0,1,2} { 
      \foreach \y in {0,1,2} {
          \draw[coupler] (lat_padding_\x\y) -- (lat_padding_\the\numexpr\x+1\relax\y);
      }
      }
        \node (labelPadding) at (1, -1.2) {\tiny Add padding};
      % vertical couplers
      \foreach \x in {0,1,2,3} {
      \foreach \y in {0,1} { 
          \draw[coupler] (lat_padding_\x\y) -- (lat_padding_\x\the\numexpr\y+1\relax);
      }
      }
      \end{scope}

      \begin{scope}[shift={(-3, -6.5)}, scale=0.5, local bounding box=map_B_step]
        \foreach \x in {0,1,2,3} \foreach \y in {0,1,2} \node[qubit] (lat_MapB_\x\y) at (\x, \y) {};
        \node[a_qubit] at (0,0) {}; \node[a_qubit] at (1,0) {}; 
        \node[p_qubit] at (0,1) {}; \node[p_qubit] at (1,1) {}; \node [p_qubit] at (2,0) {};
        \node[b_qubit] at (2, 1) {}; \node[b_qubit] at (3, 1) {};

        \node (labelAddMapB) [align=center] at (1, -1.2) {\tiny Map $M_2$,\\ \tiny Combined map is $\mathcal{M} = [M_1, M_2]$};
              % horizontal couplers
      \foreach \x in {0,1,2} { 
      \foreach \y in {0,1,2} {
          \draw[coupler] (lat_MapB_\x\y) -- (lat_MapB_\the\numexpr\x+1\relax\y);
      }
      }

      % vertical couplers
      \foreach \x in {0,1,2,3} {
      \foreach \y in {0,1} {
          \draw[coupler] (lat_MapB_\x\y) -- (lat_MapB_\x\the\numexpr\y+1\relax);
      }
      }
      \end{scope}

      % stage 2
      \node (start2) [box, font=\large] at (8.5, -0.6) {Stage 2: Execution Pipeline};

      \begin{scope}[shift={(7.2,-3.5)}]
          \node (J0) [job_base] at (0, 1.0) {Solo $\left(\mathcal{Q}_{M_1}^{C_A}\right)$};
          \node (J1) [job_base] at (0, 0.3) {Solo $\left(\mathcal{Q}_{M_1}^{C_B}\right)$};
          \node (J2) [job_zipped] at (0, -0.4) {Co $\left(\mathcal{Q}_{M_1}^{C_A} \| T_{M_2}^B\right)$};
          \node (J3) [job_zipped] at (0, -1.1) {Co $\left(\mathcal{Q}_{M_1}^{C_B} \| T_{M_2}^A\right)$};
          \node (J4) [job_base] at (0, -1.8) {Solo $\left(\mathcal{Q}_{M_1}^{C_A}\right)$};
          \node (J5) [job_base] at (0, -2.5) {Solo $\left(\mathcal{Q}_{M_1}^{C_B}\right)$};
          \draw[thick, CircuitABlue] (1.3, 1.2) -- (1.3, 0.7) node (lab1) [midway, right, font=\tiny] {Base ($C_A \text{ on } M_1$)};
          \draw[thick, CircuitABlue] (1.3, 0.5) -- (1.3, 0) node (lab1) [midway, right, font=\tiny] {Base ($C_B \text{ on } M_1$)};
          \draw[thick, CircuitBRed] (1.3, -0.2) -- (1.3, -0.7) node (lab2) [midway, right, font=\tiny] {Co $C_A \leftarrow C_B$ $(C_A \text{ on } M_1, C_B \text{ on } M_2)$};
          \draw[thick, CircuitBRed] (1.3, -0.9) -- (1.3, -1.4) node (lab2) [midway, right, font=\tiny] {Co $C_B \leftarrow C_A$ $(C_B \text{ on } M_1, C_A \text{ on } M_2)$};
          \draw[thick, CircuitABlue] (1.3, -1.6) -- (1.3, -2.1) node (lab3) [midway, right, font=\tiny] {Drift $C_A$};
          \draw[thick, CircuitABlue] (1.3, -2.3) -- (1.3, -2.8) node (lab3) [midway, right, font=\tiny] {Drift $C_B$};
      \end{scope}

      \node (drift_text) [draw=black!40, fill=white, text width=4.5cm, font=\tiny, rounded corners, inner sep=2mm] at (11.5, -1.7) {
          \textbf{Verification Strategy:}\\
          Pre/Post Solo execution (circuits are compared to how they behave on map $M_1$) used to isolate crosstalk statistics from temporal drift.
      };

      % backgrounds
      \begin{scope}[on background layer]
          \node (bg1) [fill=gray!5, draw=gray!40, dashed, rounded corners, inner sep=12pt,
                fit=(input1) (start1) (labelA) (padding_step) (bfs_text) (map_B_step)] {};
          \node[anchor=north west, font=\tiny\bfseries, text=gray] at (bg1.north west) {Stage 1: Partitioning};

          \node (bg2) [fill=gray!5, draw=gray!40, dashed, rounded corners, inner sep=12pt,
                fit=(start2) (J0) (J5) (drift_text) (lab1)] {};
          \node[anchor=north west, font=\tiny\bfseries, text=gray] at (bg2.north west) {Stage 2: Execution};
      \end{scope}

      % arrows
      \draw[arrow] (first_step.east) -- (first_step.east -| mapping_step.west);
      \draw[arrow] (mapping_step.south) -- (mapping_step.south |- padding_step.north);
      \draw[arrow] (padding_step.west) -- (padding_step.west -| map_B_step.east);
      \draw[arrow] (input1) -- (start1);
      \draw[arrow] (start1) -- (0, -2.6);
      
      % horizontal transition
      \draw[arrow, gray, dashed] (bg1.east) -- (bg1.east -| bg2.west) node[midway, above, font=\tiny, text=black] {$\mathcal{M}$ Passed};
      \draw[arrow] (start2) -- (8.5, -2.2);

  \end{tikzpicture}
  }
  \caption{Visual overview of the experimental methodology. Stage 1 executes spatial partitioning and padding to generate the mapping set $\mathcal{M}$. Stage 2 utilizes a verification pipeline (Solo vs Cotenant) to isolate crosstalk from temporal drift.}
  \label{fig:methodology_flow}
\end{figure*}
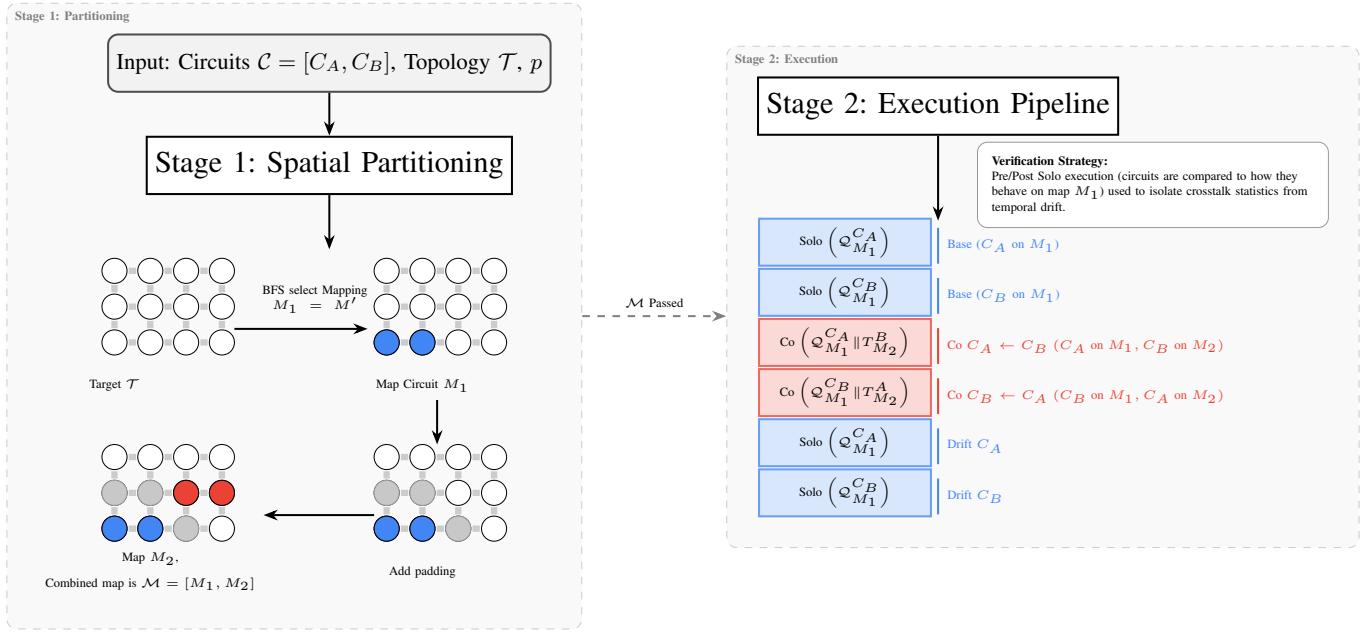

Algorithms \ref{alg:circuit_map_selection} and \ref{alg:circuit_matching} illustrate the approach of data acquisition. Crucially, the transpile steps (lines 7 and 13) impose a strict hardware constraint, forcing the Qiskit transpiler to use only the qubit subset defined by the mapping parameter $\mathcal{M}$. This containment is a prerequisite for the \texttt{transpiled\_zip} function, which mandates that the set of qubits $Q$ for any two circuits $C_i$, $C_j$ are disjoint ($Q_i \cap Q_j = \emptyset$ for $i \neq j$). Under this condition, the combined unitary operators on $\ell$ simultaneous circuits can be executed in a single quantum job without instruction overlap. Equation \eqref{eq:zipped} shows the explicit formulation for composing the global unitary operation of the independent circuits.

\begin{equation}
  U_{\mathcal{Z}} = \bigotimes_{i = 1}^{\ell} U_i(Q_i)
  \label{eq:zipped}
\end{equation}

The experimental sequence is designed to establish a robust performance baseline and account for fluctuations in hardware noise. The inclusion of isolated (standalone) circuit executions both before and after the cotenant phases functions as a control mechanism for the baseline variability of these circuits. This `sandwich' execution pattern, formalized in Alg.~\ref{alg:circuit_matching} (lines 8 and 19), facilitates the decoupling of device noise from inter-circuit interference inherent in multitenant settings. Here, ``\textbf{sandwich}" refers to the idea of having a standalone execution before and after the pairing or grouping execution. Figure~\ref{fig:methodology_flow} illustrates what this mapping and circuit execution flow would look like with two circuits on a square-lattice processor. 

To characterize the specific crosstalk interactions between circuits with high granularity, the study prioritizes pairwise circuit analysis over large-scale group execution. By restricting the cotenant setting to circuit-to-circuit pairings, we resolve specific crosstalk signatures that would be diluted in a high-load scenario. This granular approach provides a controlled environment to isolate the interference patterns between distinct circuit pairings. While multitenant environment will exhibit more complex noise patterns, this evaluation acts as a \textbf{first-order approximation} of circuit behavior in higher-load multitenant scenarios. 

To evaluate the robustness of the proposed mapping strategies, experiments were conducted across a diverse fleet of IBM Quantum processors. The standard experimental protocol consisted of 24 independent trials per backend. Each set is equally partitioned between configurations with no padding and a padding depth of $p=2$. While the majority of the fleet followed this 24-trial baseline, execution counts were adjusted for specific backends to accommodate hardware availability constraints. Specifically, \texttt{ibm\_torino} (Heron r1) was utilized for 16 trials, while \texttt{ibm\_miami} (Nighthawk r1) provided a targeted validation set of two trials. In total, 138 unique multitenant experiments were performed, providing a comprehensive dataset for characterizing crosstalk interference topographies across different noise profiles. As detailed in Table \ref{tab:machines}, the experimental fleet comprises several generations of the Heron and Nighthawk architectures, allowing for a longitudinal assessment of crosstalk across varying qubit counts and revision-specific noise profiles.

\subsection{Circuit Mapping}
To isolate the impact of spatial proximity on inter-circuit interference, the mapping process utilizes a constrained qubit selection strategy. For a given circuit pair, the mapping $M_0$ is held constant across different padding configurations, and $M_1$ is recalculated to satisfy the minimum distance requirements imposed by the padding depth $p$. This consistency ensures that observed variations in fidelity are attributable to the change in spatial isolation rather than fluctuations in the underlying qubit mappings. For each new experiment $M_0$ is recalculated to get a new baseline. All results across each machine and selected mapping is then aggregated as discussed in Section \ref{sec:quantifying_crosstalk_interference}.

As formalized in Algorithm \ref{alg:circuit_map_selection}, the selection of the frontier set $F$ (Lines 5 and 16) facilitates a `greedy' spatial allocation. By iteratively selecting mappings from the immediate neighbors of reserved qubits, the algorithm simulates a high-density scheduling scenario where circuits are packed to minimize fragmented hardware overhead. While production-level schedulers typically leverage real-time calibration data to optimize for gate fidelity, our randomized neighbor-selection approach provides a unbiased characterization of the hardware's structural interference patterns, independent of transient calibration peaks.

\subsection{Quantifying Crosstalk Interference}
\label{sec:quantifying_crosstalk_interference}
To isolate the impact of concurrent circuit execution from intrinsic hardware drift, we normalize the interference data against the standalone baseline executions by running circuits in isolation before and after pairing them. This procedure is detailed visually in Stage 2 of Figure \ref{fig:methodology_flow}. We define the output probability distribution of circuit $C_i$ (mapped to $M_0$) when executed concurrently with circuit $C_j$ (mapped to $M_1$) as $\rho_{i, j}^{(M_0, M_1)}$. To establish a control, we denote the ``sandwich'' baseline executions (where $C_i$ runs in isolation) as $\rho_{i, 0}$ (pre-stress) and $\rho_{i, m+1}$ (post-stress).

The similarity between any two execution distributions is quantified using the classical fidelity (Bhattacharyya coefficient \cite{Bhattacharyya1943}):
\begin{equation}
  F(p, q) = \sum_{x \in \{0, 1\}^n} \sqrt{p(x)q(x)},
  \label{eq:prob_dist_fidelity}
\end{equation}
where $n$ is the number of qubits in the mapping. We calculate the fidelity of each cotenant execution relative to the pre-stress baseline, $F(\rho_{i, 0}, \rho_{i, j})$. For a set of $m=5$ circuits, this yields an interference matrix where each entry represents the fidelity degradation of $C_i$ due to the presence of $C_j$.

To evaluate the relative impact of specific circuit pairings across different hardware backends, we aggregate these results over $r$ experimental trials and compute a standardized score. Let $\bar{F}(C_i, C_j)$ be the mean fidelity over $r$ trials. We then characterize the sensitivity of circuit $C_i$ by calculating the mean performance ($\mu_i$) and standard deviation ($\sigma_i$) across all cotenant scenarios, including the post-stress baseline $C_{m+1}$:

\begin{IEEEeqnarray}{rCl}
  \mu_i &=& \frac{1}{m+1}\sum_{j=1}^{m+1} \bar{F}\left(C_i, C_j\right) \label{eq:row_mean} \\
  \sigma_i &=& \sqrt{\frac{1}{m}\sum_{j=1}^{m+1} \left(\bar{F}\left(C_i, C_j\right) - \mu_i\right)^2},
  \label{eq:row_avg_std}
\end{IEEEeqnarray}
where $\mu_i$ is the average and $\sigma_i$ is the sample standard deviation of cotenant impact for the benchmarked circuits.

Finally, the Interference Impact Matrix $H^{(\mathcal{Q})}$ for a machine $\mathcal{Q}$ is defined. Each entry $H_{i, j}$ is computed as a $t$-statistic to quantify the relative deviation:
\begin{equation}
  H_{i, j}^{(\mathcal{Q})} = \frac{\bar{F}\left(C_i, C_j\right) - \mu_i}{\sigma_i}
  \label{eq:impact_matrix}
\end{equation}

This metric represents the relative impact of circuit $C_j$ on $C_i$ compared to the average interference observed for $C_i$. A negative $H_{i,j}$ indicates that circuit $C_j$ induces greater-than-average fidelity degradation, effectively identifying high-crosstalk pairings within the interference topography.

\subsection{Cross-Platform Interference Similarity}

To determine if circuit interference patterns are consistent across distinct architectural revisions, we evaluate the structural correlation between the interference impact matrices $H^{(\mathcal{Q})}$. We employ the Structural Similarity Index Measure (SSIM) to quantify this relationship. While SSIM is conventionally an image-processing metric, it is utilized here to identify the structural relationship between interference topographies across different quantum processors.

Because SSIM is defined for non-negative signals, we apply a global translation to the values in $H$. Let $g$ be the global minimum value across all matrices such that $g < 0$. We define a shifted matrix $H' = H + |g|$, effectively treating each entry as a ``pixel'' intensity representing the degree to which circuit $C_j$ (the cotenant) interferes with circuit $C_i$ (the target). As established, $H_{i,j} < 0$ indicates $C_j$ has a lower-than-average impact on $C_i$, while $H_{i,j} > 0$ indicates higher-than-average interference.

For a fleet of $k=7$ machines, we compute the pairwise similarity $\text{SSIM}(H^{\prime(\mathcal{Q}_i)}, H^{\prime(\mathcal{Q}_j)})$ using a $3 \times 3$ sliding window. This allows us to determine if specific circuit pairings exhibit universal crosstalk behaviors that transcend specific hardware revisions.

\subsection{Metrics for Circuit Aggression and Sensitivity}

Let $\mathcal{I}: \mathcal{C} \times \mathcal{C} \to \mathbb{R}$ be a function measuring the impact one circuit has on another. Where a value of $0$ indicates no discernable difference from a single-tenant scenario. To approximate this function, we isolate the impact of circuit $C_j$ on $C_i$ by subtracting the baseline single-tenant variance:
\begin{equation}
\hat{\mathcal{I}}(C_i, C_j) = H_{i, j} - H_{i, m+1}
\label{eq:approx_I}
\end{equation}
where $H_{i, m+1}$ represents the post-stress standalone baseline. 

From this pairwise impact, we derive two aggregate metrics to characterize the noise profiles of the circuit library:

\begin{enumerate}
    \item \textbf{Crosstalk Aggression ($\mathcal{N}$):} Defined as the arithmetic mean of the impact that noise emitted from circuit $C_j$ has on all other circuits $C_i$. This quantifies a circuit's tendency to generate interference as a cotenant.
    \begin{equation}
      \mathcal{N}(C_j) = \frac{1}{m} \sum_{i=1}^{m} \hat{\mathcal{I}}(C_i, C_j)
    \end{equation}
    
    \item \textbf{Crosstalk Sensitivity ($\mathcal{S}$):} Defined as the arithmetic mean of the impact all other circuits $C_j$ impose on circuit $C_i$ through crosstalk. This measures how susceptible a circuit is to multitenant execution.
    \begin{equation}
      \mathcal{S}(C_i) = \frac{1}{m} \sum_{j = 1}^{m} \hat{\mathcal{I}}(C_i, C_j)
    \end{equation}
\end{enumerate}

\subsection{Scheduling Objective and Optimization Target}

Rather than proposing a specific heuristic, we define the optimization target for a crosstalk-aware scheduler. Given a lookup table of circuit types and their empirically determined crosstalk affinities $\hat{\mathcal{I}}$, the objective is to minimize the aggregate expected crosstalk across the processor, ensuring that each constituent circuit maintains maximum fidelity in the multitenant setting.

Let $\mathcal{C}$ represent a queue of pending circuits managed by a priority scheduler. The scheduler selects a subset of circuits, $\mathcal{C}_{\mathcal{Q}} \subseteq \mathcal{C}$, to be executed concurrently on the processor $\mathcal{Q}$. The goal of the scheduler is to minimize the total inter-circuit interference within the multitenant workload, formulated as the following objective function:

\begin{equation}
  \argmin_{\mathcal{C}_{\mathcal{Q}}} \sum_{(C, C') \in \mathcal{C}_{\mathcal{Q}} \times \mathcal{C}_{\mathcal{Q}}} \hat{\mathcal{I}}(C, C')
  \label{eq:objective_function}
\end{equation}

By minimizing this sum, the scheduler identifies the most compatible circuit combinations based on the interference topographies characterized in Section \ref{sec:quantifying_crosstalk_interference}. This formulation serves as the theoretical basis for any subsequent scheduling algorithms, such as greedy placement or simulated annealing, that utilize our characterized crosstalk data.

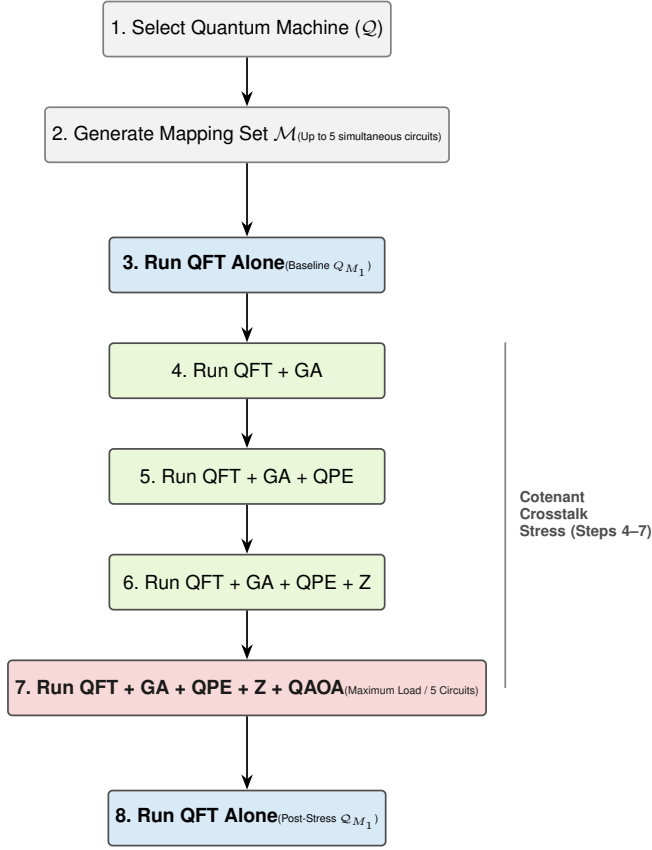
\begin{figure}
  \resizebox{\columnwidth}{!} {
\begin{tikzpicture}[
    node distance = 0.8cm and 1.0cm,
    every node/.style={font=\small\sffamily},
    box/.style = {rectangle, draw=black, fill=white, minimum width=4.5cm, minimum height=0.9cm, text centered, thick, rounded corners=2pt},
    setup/.style = {box, fill=SetupGray, draw=gray},
    baseline/.style = {box, fill=BaselineBlue, draw=black!70},
    stress/.style = {box, fill=StressGreen, draw=black!70},
    maxstress/.style = {box, fill=MaxStressRed, draw=black!70, text=black!90},
    arrow/.style = {thick, -{Stealth[length=2.5mm]}, black},
    dashed_arrow/.style = {thick, dashed, -{Stealth[length=2.5mm]}, gray}
]

    \node (machine) [setup] {1. Select Quantum Machine ($\mathcal{Q}$)};
    
    \node (mapping) [setup, below=of machine] {2. Generate Mapping Set $\mathcal{M}$ \\ \tiny (Up to 5 simultaneous circuits)};

    \node (pre_alone) [baseline, below=1.2cm of mapping] {\textbf{3. Run QFT Alone} \\ \tiny (Baseline $Q_{M_1}$)};

    \node (ga) [stress, below=of pre_alone] {4. Run QFT + GA};
    
    \node (qpe) [stress, below=of ga] {5. Run QFT + GA + QPE};
    
    \node (z) [stress, below=of qpe] {6. Run QFT + GA + QPE + Z};
    
    \node (qaoa) [maxstress, below=of z] {\textbf{7. Run QFT + GA + QPE + Z + QAOA} \\ \tiny (Maximum Load / 5 Circuits)};

    \node (post_alone) [baseline, below=1.2cm of qaoa] {\textbf{8. Run QFT Alone} \\ \tiny (Post-Stress $\mathcal{Q}_{M_1}$)};

    \draw [arrow] (machine) -- (mapping);

    \draw [arrow] (mapping) -- (pre_alone);

    \draw [arrow] (pre_alone) -- (ga);
    \draw [arrow] (ga) -- (qpe);
    \draw [arrow] (qpe) -- (z);
    \draw [arrow] (z) -- (qaoa);

    \draw [arrow] (qaoa) -- (post_alone);

    \draw[thick, black!50] 
    ([xshift=0.3cm]qaoa.east |- ga.north) -- 
    ([xshift=0.3cm]qaoa.east)
    node [midway, right=3pt, text width=2.2cm, align=left, font=\scriptsize\sffamily\bfseries, text=black!70] {
        Cotenant Crosstalk\\ Stress (Steps 4--7)};

    \path (pre_alone.west) --++ (-0.8cm,0) coordinate (pre_anchor);
    \path (post_alone.west) --++ (-0.8cm,0) coordinate (post_anchor);

\end{tikzpicture}
  }
  \caption{Flow of adding more and more stress on QFT circuit multitenancy with the strong to weak interference circuits. The weak to strong interference add the circuits in reverse order starting with adding QAOA and ending with adding GA.}
  \label{fig:incremental_stress}
\end{figure}

\subsection{Scaling Analysis: Cumulative Crosstalk Stress}

To evaluate the validity of the pairwise approximation in high-load scenarios, we perform a cumulative stress-test by incrementally increasing the number of concurrent circuits. Using the Quantum Fourier Transform (QFT) as the target circuit, we execute a series of multitenant workloads ranging from a single-tenant baseline to a maximum load of five concurrent circuits.

As illustrated in Figure \ref{fig:incremental_stress}, we employ the sandwiched execution protocol to account for temporal drift on the QFT circuit. We evaluate the cumulative impact using two distinct priority sequences based on the previously characterized interference topographies: 
\begin{enumerate}
    \item \textbf{Strong-to-Weak (STW):} Circuits are added in descending order of their crosstalk aggression ($\mathcal{N}$), prioritizing the high-interference pairings first. Figure \ref{fig:incremental_stress} illustrates the strong-to-weak pattern of the experimental design.
    \item \textbf{Weak-to-Strong (WTS):} Circuits are added in ascending order of their crosstalk aggression against QFT, incrementally increasing the noise floor.
\end{enumerate}

This dual-directional approach allows us to determine if the order of allocation—and the resulting spatial adjacency—non-linearly impacts the fidelity of the target circuit. By comparing the STW and WTS sequences.

\begin{figure}[htbp]
  \includegraphics[width=\columnwidth]{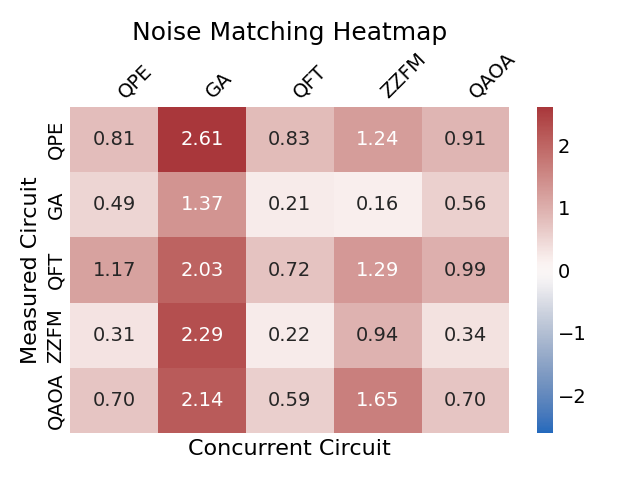}
  \caption{Overall interaction scores between circuits across all machines. All noise levels have been shifted so that the standalone baseline is 0. The columns show an uneven structure indicating that circuits are not simply "noisy" but convey a complex relationship.}
  \label{fig:standalone_controlled}
\end{figure}

\section{Results}

\begin{figure}[htbp]
  \includegraphics[width=\columnwidth]{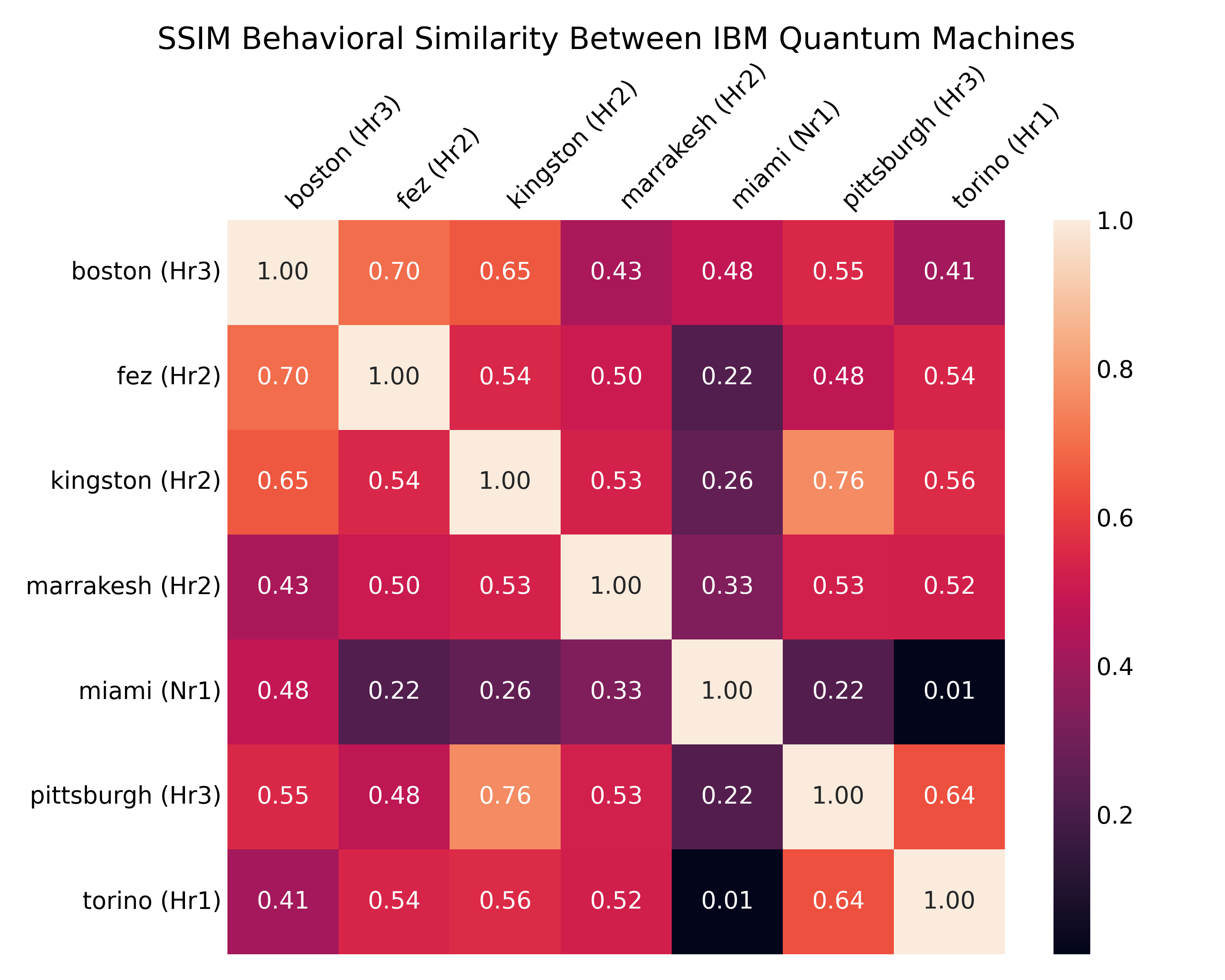}
  \caption{This chart illustrates how similarly each of the machines behave. A higher score means that a scheduler calibrated for one machine can be deployed for another machine. The sequence of lower scores for \texttt{ibm\_miami} (Nighthawk r1) indicates a recalibration for complete revisions of new circuits will be needed.}
  \label{fig:machine_similarity}
\end{figure}

\begin{figure}[htbp]
  \includegraphics[width=\columnwidth]{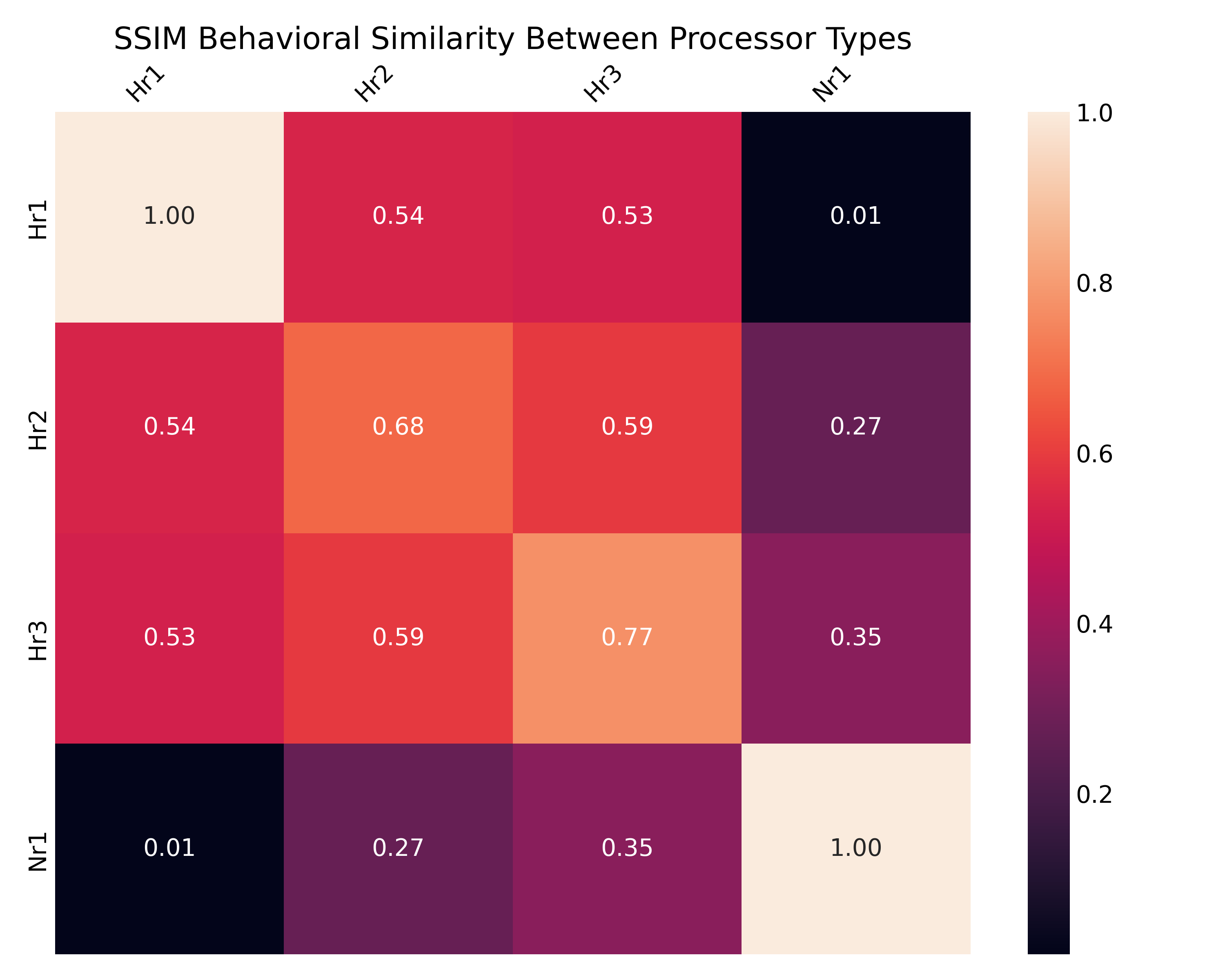}
  \caption{A higher level overview of circuit types. These values are averaged amount the SSIM scores between each processor type. We see that there is a noticeable drop in the score between the Nighthawk r1 and Heron r1 processors where the score is approximately 0, indicating a need for a complete scheduler recalibration across generations. However, only a slight recalibration would be needed for processors of the same type.}
  \label{fig:processor_similarity}
\end{figure}

This work evaluates the importance of circuit selection in multitenancy. Our initial results quantify the interference that each circuit has on each other on a variety of machines. Our evaluation determines there is a \textbf{complex topography of circuit interference patterns from crosstalk in multitenant scenarios}, illustrating the idea that circuits are not simply noisy or sensitive, but the circuit pairing interaction is more nuanced. Such a nuance suggests that a scheduler making the decision on which circuits to run concurrently on a quantum computer must consider how these circuits will interact with each other. Afterwards, we show this \textbf{behavior is consistent across machines under certain conditions, indicating when the scheduler can transfer its calibration to another machine}. Consistency allows the same scheduler to learn the behavior of circuit to circuit and transfer knowledge to another machine. Finally, we \textbf{show that the spatial mapping of the circuits has an influence on the outcome} by evaluating the placement order of strong to weak influence and weak to strong influence for the target evaluation circuit.

\begin{figure*}[htbp]
  \centering

  \begin{subfigure}[b]{0.48\textwidth}
      \centering
      \includegraphics[width=\textwidth]{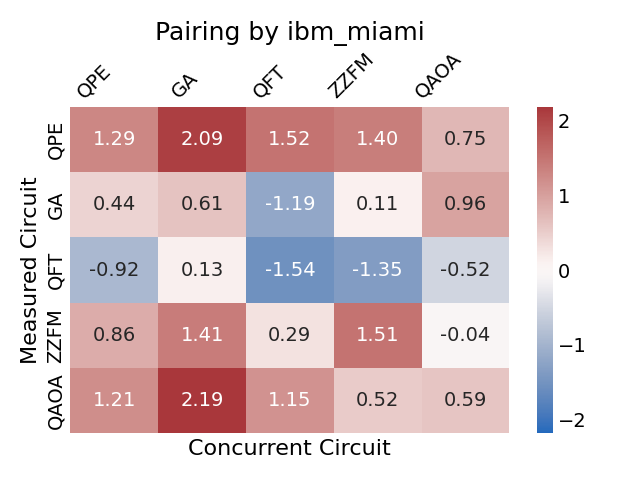}
      \caption{Heatmap of \texttt{ibm\_miami} (Nighthawk r1) interactions.}
      \label{fig:plot_one}
  \end{subfigure}
  \hfill 
  \begin{subfigure}[b]{0.48\textwidth}
      \centering
      \includegraphics[width=\textwidth]{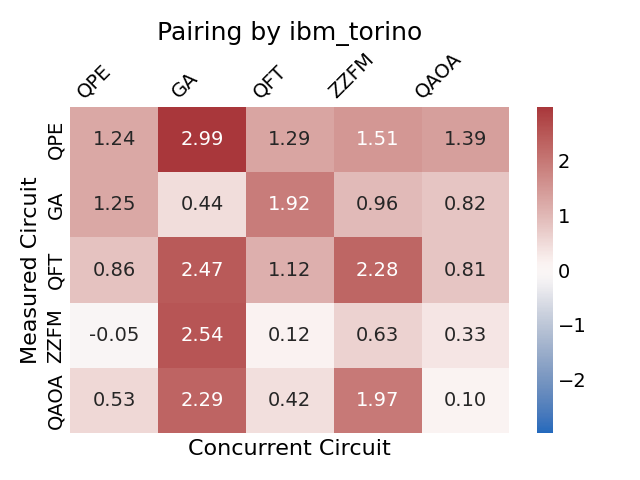}
      \caption{Heatmap of \texttt{ibm\_torino} (Heron r1) circuit interactions.}
      \label{fig:plot_biggest_diff}
  \end{subfigure}
  
  \caption{\texttt{ibm\_torino} and \texttt{ibm\_miami}, the two processors with the lowest SSIM score. We can see that some interactions on \texttt{ibm\_miami} are negative. This is an indication that the impacting circuit crosstalk was statistically indistinguishable from standard machine variation. For \texttt{ibm\_miami} QFT was the least sensitive circuit. Because SSIM expects all positive values, as with all matrices, the comparison was done with a static shift up by $|g|$. Here $g$ represents the global minimum score for any circuit pairing.}
  \label{fig:extreme_diff}
\end{figure*}

Figure \ref{fig:standalone_controlled} illustrates the behavior between circuits across all machines. The results show that these circuits display a complex dynamic relation between what circuits are on the machine. We find that for example, QPE and QFT present a low-interference pattern except against QFT. Circuits like ZZFM are robust except against ZZFM and Grover's algorithm, displaying very little differentiation against the standalone scenario. Using this information we can categorize four circuit types highly noisy, low interference, highly sensitive, and robust. Quantitatively, circuits like QFT exhibit a sensitivity score of $1.24>1$ indicating high sensitivity to cotenant execution, whereas ZZFM shows sensitivity $0.82<1$. When excluding the impact of Grover's the noisiest circuit the sensitivity scores of QFT and ZZFM change to $1.04>1$ and $0.45<0.5$ respectively. This shift indicates that ZZFM is significantly less influenced by the crosstalk noise than QFT.

\begin{figure*}
  \includegraphics[width=\textwidth]{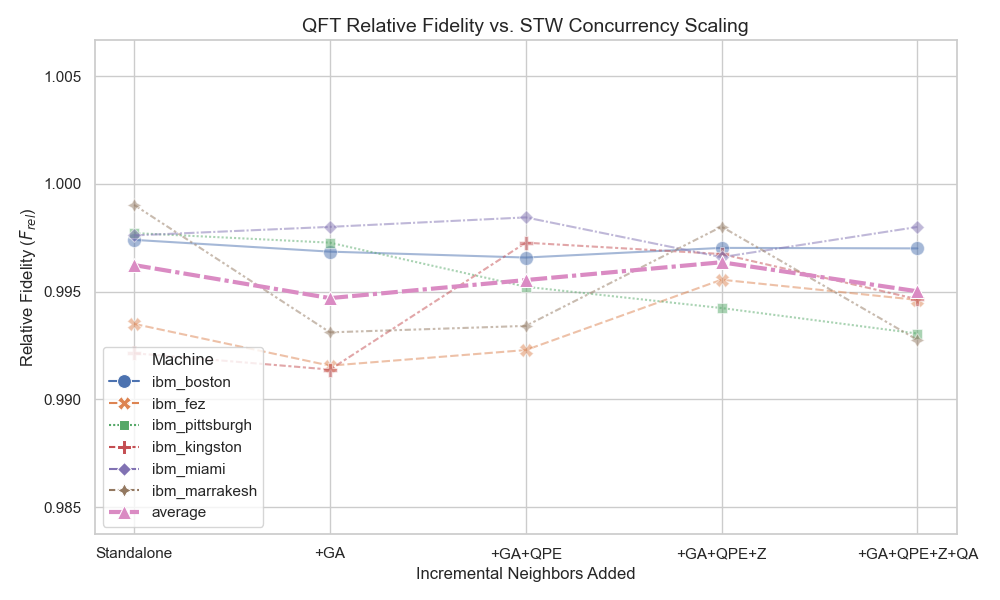}
  \caption{The strong to weak incremental addition of circuits. There is a statistically insignificant slope, and the p-value measuring non-zero slope is $0.799 > 0.05$ indicating that the strong to weak pattern does not hold a consistent result.}
  \label{fig:add_circuits_stw}
\end{figure*}

To determine how these circuits matchups remain consistent across machine-platforms, we evaluate Figure \ref{fig:machine_similarity}. This figure shows the SSIM score between each machine. We see that the SSIM score is strongly correlated for most of the machines. High scores between two machines indicate that a scheduler can be adapted without needing to recalibrate. However, one key point is that \texttt{ibm\_miami} (Nighthawk r1) has low scores for all machines with the highest being with \texttt{ibm\_boston} (Heron r3). These SSIM scores between \texttt{ibm\_miami} and the other machines is to be expected, due to the significant topological difference (moving from Heavy-Hex lattice in the Heron r3 processor to square lattice in the Nighthawk processor) in the qubit coupling. This indicates that a scheduler calibrated for one quantum processor series will need to recalibrate for another.

\begin{figure*}
  \includegraphics[width=\textwidth]{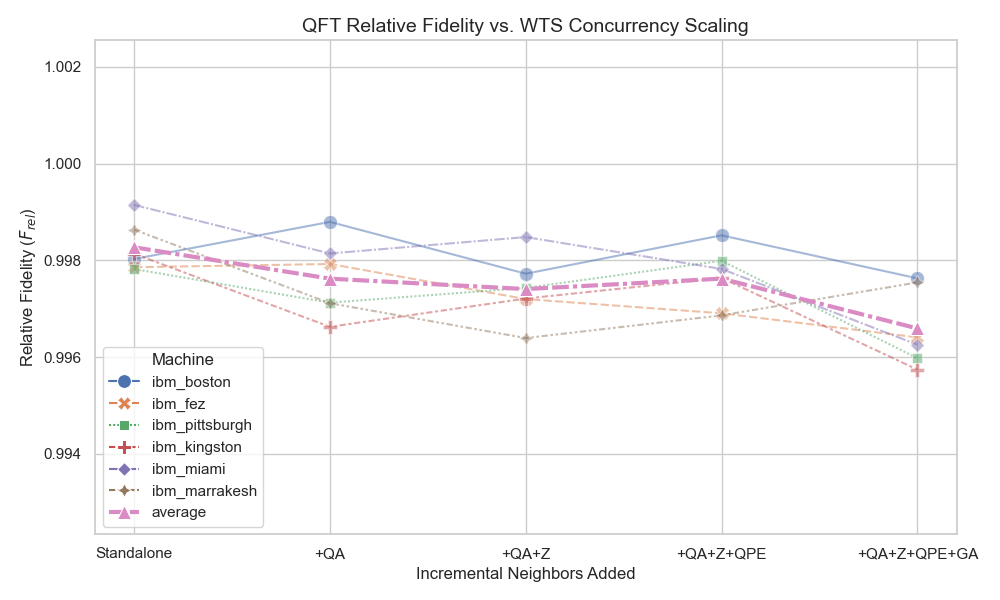}
  \caption{The weak to strong incremental addition of circuits. Points are collected across all machines. There is a slightly negative slope and p-value $0.007<0.01$. Indicating a strong correlation to more circuits leading to more noise.}
  \label{fig:add_circuits_wts}
\end{figure*}

We group the machine types into their specific revisions, and evaluate their average SSIM score to assess how strongly this connection between revisions are, and to determine how well a schedule calibrated for one processor type can perform on another processor type. Figure \ref{fig:processor_similarity} illustrates these differences.

We compare the heatmaps of the maximally different processors in our dataset to get an idea of how differently these two processors behave. 
Figure \ref{fig:plot_biggest_diff} shows the heatmap interactions between \texttt{ibm\_torino} and \texttt{ibm\_miami}. The score difference was calculated by a shifted version of both of these matrices (the shift amount $g$ being equal for all matrices). One may notice that QFT has a positive $t$ score on \texttt{ibm\_miami} for the QFT score. This is noticeably different from the behavior of the Heron processors where QFT was the most stable circuit in output deviation.

\subsection{Adding More Circuits}

To understand how circuit selection placement impacts the outcome of the circuits, we evaluate the evolution of adding circuits from STW and WTS patterns for the circuits against the QFT circuit. Figures \ref{fig:add_circuits_stw} and \ref{fig:add_circuits_wts} illustrate the overall noise patterns we see in the circuits across all machines. We see that the impact of WTS has higher impact. With a slope of $-0.0003$ and a p-value of $0.007<0.01$ indicating a definitive though small increase in error as more circuits are added. However, the STW adding pattern did not show any noticeable impact to the overall pattern. Indicating that the placement of circuits on the machine has an impact on how much the extra circuits will interfere with one given circuit.

Despite being the newest machine, \texttt{ibm\_miami} (Nighthawk r1) shows a strong downward trend on the fidelity match as more circuits are added to the concurrent execution. With a difference of $0.0029$ between the standalone case and the 5 cases, it has the largest range of difference out of all of the machines. This indicates that newer doesn't mean better. This critical difference between the Nighthawk and Heron processors indicate that the square lattice may be causing more crosstalk noise, indicating a need for more careful consideration when deciding to move towards multitenant execution.

\section{Discussion and Future Work}

We evaluated the level of impact each circuit type has on one another. The data collected here can be utilized by a circuit scheduler that must decide what quantum circuits will be run on the processor concurrently. 

The scheduler can take advantage of knowledge on four different types of circuits, and their intricate interaction patterns to decide on which circuits to run concurrently. These include universally aggressive circuits which cause a high impact to any of the other circuits, circuits which are universally sensitive (low variability in the single tenant case compared to the other cases), and circuits which highly impact or are highly sensitive to only a subset of circuits. The sensitive circuits are strong candidates for circuit snooping, these circuits can be utilized to snoop the activity of unsuspecting victim users due to their variability in their output in a multitenant setting. This adversarial behavior can be mitigated through careful scheduling and well calibrated systems. The aggressive circuits are candidates for fault-injection attacks. The adversary's attacks would be thwarted by a scheduler that can catch these from identifying the circuits. Aggressive circuits can intentionally impact the behavior of the other circuits on the processor. A scheduler which can identify the difference between these types of circuits can make isolation decisions on circuits which pose a high risk for the circuits already picked for execution on the processor.

The portability of a machine evaluation for circuit level comparison depends on how similar machines are in their crosstalk behavior. Our evaluation results show that there is some reliable translation between machines of the same type. This will allow a scheduler to learn the behavior of a machine family without having to test every additional machine individually. The lack of similarity between Nighthawk and Heron suggests that the topological differences between the processors drastically changes the crosstalk behavior and large changes in the architecture would require the scheduler to recalibrate circuit pairing affinities.

\paragraph*{Future Work} Further evaluation would have to look at random, deeper circuits and see how these circuits impact outcomes of each other. These random circuits could then be used to train a model to predict the level of affinity (cotenant capatibility) between circuits based on the circuit structure. 

The values of the fidelity measure from the original circuit remain high despite the crosstalk variance. This small crosstalk variance increases as the complexity of the circuits increase. 

\paragraph*{Limitations} While this study provides a foundational understanding of circuit interaction in a multitenant setting, there are several constraints that define the scope of the findings:
\begin{itemize}
  \item \textbf{Circuit Scale and Diversity:} Circuit selection is small, a more robust study will include a larger amount of circuits.
  \item \textbf{Topological Comparison:} Limited hardware availability prevents us from identifying further the differences between the square-lattice and the Heavy-Hex lattice architectures. There is also no data on how similarly multiple square lattice processors behave. 
  \item \textbf{Experimental repetition and drift:} The experiments repeated the exact circuits on the same qubits several times. Because the data does not include results from several consecutive single standalone runs, we do not account for ``wear" on the qubits from overuse. 
\end{itemize}

\section{Acknowledgements}

The Mizzou Quantum Innovation Center provides the resources to perform execution on the quantum computers, and financial support to Andrew Woods.

Andrew Woods has previously been funded by the National Science Foundation DEG-1946619.
\nocite{*}
\printbibliography
\end{document}